\documentclass[onecolumn, letter]{emulateapj}

\usepackage{amsmath, amssymb}

\slugcomment{Submitted to Astrophys. J.}

\shorttitle{Hanle effect for dipoles}
\shortauthors{Manso Sainz and Mart\'\i nez Gonz\'alez}

\begin{document}

\title{Hanle effect for stellar dipoles and quadrupoles}

\author{R. Manso Sainz and M. J. Mart\' inez Gonz\'alez}
\affil{Instituto de Astrof\' isica de Canarias, V\' ia L\'actea s/n, E-38205 La Laguna, Tenerife, Spain}
\affil{Departamento de Astrof\'\i sica, Universidad de La Laguna, Tenerife, Spain}

\begin{abstract}
We derive exact expressions for the degree of lineal polarization over a resolved or integrated stellar disc due to resonance scattering and the Hanle effect from a dipolar or quadrupolar distribution of magnetic fields.
We apply the theory of scattering polarization within the formalism of the spherical tensors representation for the density matrix and radiation field.
The distribution of linear polarization over the stellar disk for different configurations of the magnetic field is studied and its topology discussed.
For an unresolved dipole, the resulting polarization can be expressed in terms of just three functions (of the inclination angle and effective dipole strength), that are calculated numerically and their behaviour discussed.
Dipolar and (aligned) quadrupoles are considered in some detail, but the techniques here ---in particular, the extensive use of the spherical tensor formalism for polarization---, can easily be applied to more general field configurations.
\end{abstract}


\maketitle

\section{Introduction}

The scattering of light in a plasma produces linear polarization in spectral lines and continuum.
As a consequence, the spectrum from a stellar atmosphere is linearly polarized towards the stellar limb.
A remarkable example of this is the linearly polarized component of the Fraunhofer spectrum observed close to the solar limb, also called {\em second solar spectrum} for its radically different nature and structure with respect to the intensity spectrum \citep[][]{StenfloTwerenboldEtal83, StenfloKeller97, Gandorfer00, Gandorfer02, Gandorfer05}.
In other (unresolved) stars, its observation is difficult because the scattering polarization spectrum cancels when averagedd over the whole stellar disk, its observation requiring imperfect cancellations due to, for example, eclipses or transits \citep{Landi+88, CarciofiMagalhaes05}.

The presence of a weak magnetic field partially removes the degeneracy of the magnetic sublevels of the scatterers, which perturbs the scattering process and hence, the polarization pattern (Hanle effect) \citep{MoruzziStrumia91}.
By {\em weak} we mean that $0.1\lesssim\nu_L\tau\lesssim 10$, where the $\nu_L=1.3996\times 10^6 B$ is the Larmor frequency (expressed in s$^{-1}$) corresponding to the magnetic field $B$ (in G), and $\tau$ the characteristic time for the scattering process ---for resonance scattering with unpolarized lower level, $\tau\sim 1/A_{u\ell}$, $A_{u\ell}$ being the Einstein coefficient for spontaneous emission in the transition\footnote{This is just the slowest timescale in the absorption-reemission process. 
The faster absorption time scale $\tau_{\rm abs}=1/B_{\ell u}J$, with $B_{\ell u}$ the Einstein coefficient for absorption and $J$ the mean intensity, is irrelevant here since the lower level will be assumed to be unpolarized (Section~2), thus playing no role in the polarization of the scattered light.}.
Typically, this is a regime of fields weaker than the characteristic regime for the Zeeman effect ($\nu_L\gtrsim\Delta_D\nu$; $\Delta_D\nu$ being the Doppler width of the spectral line). 
The possibility of exploiting the Hanle effect for diagnosing solar magnetic fields in magnetic field regimes and topological configurations not easily accessible through classical Zeeman techniques is one of the reasons why the second solar spectrum has been so actively researched in recent years \citep[see][and references therein]{Stenflo82, CasiniLandi08, Trujillo09, Stenflo09}.
Moreover, the possibility has arisen that the Hanle effect could also be applied to diagnose stellar magnetic fields \citep[][and references therein]{IgnaceNordsieckEA95, IgnaceCassinelliEA97, Nordsieck01, Ignace01}.

The presence of a global magnetic field over the stellar surface brakes the symmetry limitations of the pure scattering.
Hanle effect signals depend on the orientation of the magnetic field with respect to the line of sight (LOS) {\em and} on the geometry of the scattering event (i.e., the relative orientation of the magnetic field with respect to the stellar surface).
It is then possible that scattering polarization signals appear even at disk center (for inclined magnetic fields there) and not only towards the stellar limb, and perfect cancellation when averaging over the stellar disk is extremely unlikely (but for very special field configurations).
Here, we derive exact expressions for the intensity and polarization produced by scattering polarization and the Hanle effect in the presence of a poloidal distribution of fields \citep[see also][]{LopezAristeAsensioRamosEA11, IgnaceHoleEA11}. 
We will consider with some detail relatively simple distributions of the magnetic field (a dipole, an aligned quadrupole) over the stellar disk, though more general configurations could be studied along the same lines.
Stellar dipoles have been used to study the polarization due to the longitudinal Zeeman effect from stars \citep[e.g.,][]{Schwarzschild50}, the transversal Zeeman effect \citep[e.g.,][]{LandolfiEtAl93}, synchrotron radiation from stars \citep[][]{Thorne63} and planets \citep[][]{Chang62}.

The next section summarizes theory of the Hanle effect in a resonance transition, within the spherical tensor components formalism of the density matrix.
The theory is then applied (Section 3) to calculate the linear polarization emission in a stellar disk with a poloidal distribution of fields, for which extensive use of the spherical tensors formalism is done.
The distribution of polarization on a resolved disk is discussed in Section 4; the polarized emission by an unresolved oblique rotator in Section 5.
The discussion is extended to quadrupolar fields in Section 6.

\section{Scattering Line Polarization and Hanle Effect}

\begin{figure}
\centering\includegraphics[width=17cm]{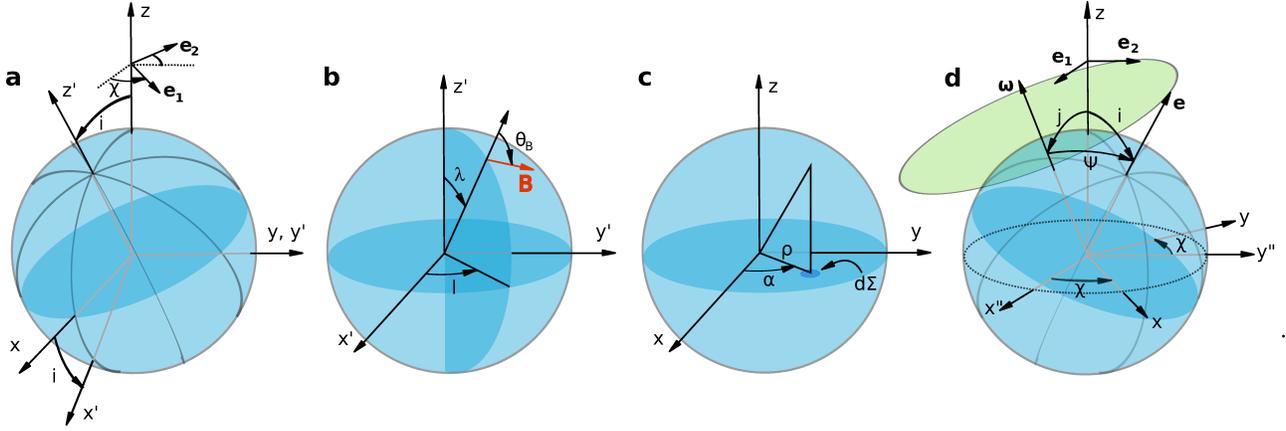}
\caption{(a) We consider a Cartesian reference system with the $z$ axis along the line-of-sight (LOS), and the stellar dipole lying in the $x$-$z$ plane, at an angle $i$ with respect to the LOS. 
The reference polarization direction $\boldsymbol{e}_1$ (positive direction for Stokes-$Q$) forms an arbitrary angle $\chi$ with the $x$ axis; correspondingly, $\boldsymbol{e}_2$ is at an angle $\chi$ with respect to the $y$ axis.
(b) In the stellar reference system, $z'$ lies along the dipole, and $y'=y$. 
At an arbitrary point on the stellar surface with longitude $l$ from the $x'$-$z'$ plane, and colatitude $\lambda$, the magnetic field $\boldsymbol{B}$ lies in the meridian plane (poloidal field), and is inclined an angle $\theta_B$ with respect to the vertical. 
(c) Averaging over the projected stellar disk can be done in a number of different ways:
$\iint{\rm d}\Sigma=\int_{-R}^{R}\int_{-\sqrt{R^2-x^2}}^{\sqrt{R^2-x^2}}{\rm d}x{\rm d}y
=\int_0^R\int_0^{2\pi}\rho{\rm d}\rho{\rm d}\alpha$
(d) In an oblique rotator, the magnetic axis $\boldsymbol{e}$ is inclined at an angle $\Psi$ with respect to the rotational axis $\boldsymbol{\omega}$ which, in turn, is at an angle $j$ to the LOS.
We choose the reference direction $\boldsymbol{e}_1$ for polarization, along the projected rotation axis (i.e., in the $x''$-$z$ plane formed by $\boldsymbol{\omega}$ and the LOS). 
}
\end{figure}

Neglecting the Zeeman splitting between $\sigma$ and $\pi$ components of the line profile (Hanle effect regime),
the emissivity in the Stokes parameters $I$, $Q$, and $U$ along a given line-of-sight (LOS; see Figure 1a) in a transition with a polarized upper level can be expressed as \citep[][hereafter LL04]{Landi+90,LandiLandolfi04}:
\begin{align}
\epsilon^{\rm line}_I&=\epsilon_0 (\rho^0_0+w^{(2)}_{J_uJ_\ell}\frac{1}{\sqrt{2}}\rho^2_0),\label{eq00} \\
\epsilon^{\rm line}_Q&=-\epsilon_0 w^{(2)}_{J_uJ_\ell}\sqrt{3}(\cos 2\chi \tilde{\rho}^2_2
-\sin 2\chi \hat{\rho}^2_2), \label{eq000}\\
\epsilon^{\rm line}_U&=\epsilon_0 w^{(2)}_{J_uJ_\ell}\sqrt{3} (\sin 2\chi \tilde{\rho}^2_2
+\cos 2\chi \hat{\rho}^2_2),\label{eq0000}
\end{align}
where $\epsilon_0=\frac{h\nu}{4\pi}A_{u\ell}N\sqrt{2J_u+1}\phi_\nu$ ($A_{u\ell}$
is the Einstein coefficient for spontaneous emission, $N$ the number density of atoms,
and $\phi_\nu$ the absorption profile), $w^{(2)}_{J_uJ_\ell}$ is
a numerical factor depending on the total angular momentum $J$ of the levels
involved in the transition (see \cite{LandiLandolfi04} for explicit values), 
and $\rho^K_Q$ ($K=0, ..., 2J$; $Q=-K, ..., K$) are the spherical components 
of the density matrix \citep{Fano57}
of the upper level of the transition (defined here taking the quantization axis along the LOS), 
which can be expressed explicitly in terms of its real ($\tilde{\rho}^K_Q$) and imaginary ($\hat{\rho}^K_Q$) parts: $\rho^K_Q=\tilde{\rho}^K_Q+{\rm i}\hat{\rho}^K_Q$.
(In particular, $\rho^0_0$ is $1/\sqrt{2J_u+1}$ times the relative population of the upper level; $\rho^2_0$ a measure of the imbalance of populations between the magnetic sublevels with a well defined projection $M_J$ of the angular momentum on the LOS).

In the magnetic field reference system (quantization axis along the magnetic 
field), the density matrix satisfies \citep[LL04]{Landi+90}:
\begin{equation}\label{eq02}
[1+\delta^{(K)}(1-\epsilon)]\frac{2h\nu^3}{c^2}\frac{2J_\ell +1}{\sqrt{2J_u+1}} \rho^K_Q=
\frac{1}{1+{\rm i}\Gamma Q}(1-\epsilon)w^{(K)}_{J_uJ_\ell} (-1)^Q \bar{J}^K_{-Q}
+\epsilon B_\nu \delta_{K0}\delta_{Q0},
\end{equation}
where $\delta^{(K)}=D^{(K)}/A_{u\ell}$, $D^{(K)}$ is the elastic collisions depolarizing rate, 
$\epsilon=C_{u\ell}/(A_{u\ell}+C_{u\ell})$, $C_{u\ell}$ is the collisional de-excitation rate, 
$\Gamma=\gamma(1-\epsilon)/[1+\delta^{K}(1-\epsilon)]$, $\gamma=2\pi\nu_L g_{u}/A_{u\ell}$,
$\nu_L=1.3996\times 10^6 B$ is the Larmor frequency in s$^{-1}$, $B$ the magnetic field in G, 
$g_u$ the Land\'e factor of the upper level, and $B_\nu$ is the Planck function.
The radiation field tensor spherical components $J^K_Q(\nu)$ ($K=0, 1, 2$; $Q=-K, ..., K$) are averages over all directions of the radiation field illumination incident upon a point \citep[][LL04]{Landi84}.
In particular, $J^0_0(\nu)=\oint\frac{{\rm d}\Omega}{4\pi}I_{\mu\chi'}$ (${\rm d}\Omega=\sin\theta{\rm d}\theta{\rm d}\chi'={\rm d}\mu{\rm d}\chi'$ is the element of solid angle), $J^1_Q(\nu)=\sqrt{\frac{3}{2}}\oint\frac{{\rm d}\Omega}{4\pi}V_{\mu\chi'}$, and
$J^2_0(\nu)=\oint\frac{{\rm d}\Omega}{4\pi}\frac{1}{2\sqrt{2}}[(3\mu^2-1)I_{\mu\chi'}+3(\mu^2-1)Q_{\mu\chi'}]$ are the only non-vanishing components if the radiation field is axially-symmetric and independent of $\chi'$ (e.g., in a plane-parallel medium or in a spherically symmetric atmosphere).
The radiation field tensors are averaged through the absorption profile in Equation~(\ref{eq02}), $\bar{J}^K_Q=\int{\rm d}\nu\phi_\nu J^K_Q(\nu)$, in accord with the complete redistribution hypothesis (the frequency of the incident photon is {\em redistributed}, collisionally or otherwise, and unrelated to that of the incident scattered photon). 
Hence, $\bar{J}^1_Q\equiv 0$, since  there is no net circular polarization in the medium.

\section{Disk distribution and integration over the stellar disk}


A rotation $R$ (Euler angles $(\alpha\beta\gamma)$) from an ``old'' reference system to a ``new'' one transforms 
the spherical components of the density matrix (the ``*'' symbol stands for complex conjugate), and radiation field
tensors according to the following rules
$$[\rho^K_Q]^{\rm new}=\sum_{Q'}[\rho^K_{Q'}]^{\rm old}{\cal D}^K_{Q'Q}(R)^*, \qquad\qquad
[J^K_Q]^{\rm new}=\sum_{Q'}[J^K_{Q'}]^{\rm old}{\cal D}^K_{Q'Q}(R),$$
where ${\cal D}^K_{PQ}(\alpha\beta\gamma)=d^K_{PQ}(\beta){\rm e}^{-{\rm i}(P\alpha+Q\gamma)}$
is the rotation matrix and $d^K_{PQ}$ the reduced rotation matrix \citep{BrinkSatchler68}.
Hence, $J^K_{-Q}=\sum_q[J^K_{-q}]_v d^K_{-q-Q}(\theta_B)$ in Equation~(\ref{eq02}), where 
$[J^2_{-q}]_v$ are the values of the radiation field tensors in a reference system with the
quantization axis along the local vertical direction, and $\theta_B$ is
the inclination of the magnetic field with respect to the local vertical (Fig.~1b).
Multiplying Equation~(\ref{eq02}) by ${\cal D}^2_{QQ'}(0, -\Theta_B, -\ell)^*$ and 
${\cal D}^2_{Q'Q''}(0, -i, 0)^*$ ($\Theta_B=\theta_B+\lambda$, $\lambda$ is 
the colatitude with respect to the dipole; see Figure~1b), 
and summing over $Q$ and $Q'$, we get the density matrix in the reference system 
with the quantization axis along the LOS and the dipole vector contained in the $x$-$z$ plane. 
After some index renaming:
\begin{multline}\label{eq06}
[1+\delta^{(K)}(1-\epsilon)]\frac{2h\nu^3}{c^2}\frac{2J_\ell +1}{\sqrt{2J_u+1}} \rho^K_Q=
\\
(1-\epsilon)w^{(K)}_{J_uJ_\ell} \sum_q (-1)^q [\bar{J}^K_{-q}]_v \sum_{Q'Q"}\frac{1}{1+{\rm i}\Gamma Q'}
d^K_{qQ'}(\theta_B)d^K_{Q'Q"}(-\Theta_B)d^K_{Q"Q}(-i){\rm e}^{-{\rm i}\ell Q"} \\
+\epsilon B_\nu \delta_{K0}\delta_{Q0}.
\end{multline}
Equation~(\ref{eq06}) is in the $xyz$ reference system (see Figure~1), and can thus be directly plugged into Equations~(\ref{eq00})-(\ref{eq0000}).

Computing the exact $\bar{J}^K_Q$ components at all points in a plane-parallel, spherical, or multidimensional atmosphere is a challenging numerical problem \cite[e.g.,][]{IgnaceCassinelliEtal99, TrujilloManso99, Nagendra03}.
In an optically thick atmosphere, $J^0_0\gg J^2_0 \gg |J^2_{Q\ne 0}|$, which means that the radiation field is only weakly anisotropic and that azimuthal symmetry breaking effects (due to horizontal inhomogeneities or magnetic fields) are negligible \citep{Manso02, MansoTrujillo11}. 
Thus, neglecting $\bar{J}^2_{Q\ne 0}$ elements in Equation~(\ref{eq06}), for $\rho^2_Q$:
\begin{equation}\label{eq07}
[1+\delta^{(K)}(1-\epsilon)]\frac{2h\nu^3}{c^2}\frac{2J_\ell +1}{\sqrt{2J_u+1}} \rho^2_Q=
(1-\epsilon)w^{(2)}_{J_uJ_\ell} [\bar{J}^2_0]_v \sum_{Q'Q"}\frac{1}{1+{\rm i}\Gamma Q'}
d^2_{0Q'}(\theta_B)d^2_{Q'Q"}(-\Theta_B)d^2_{Q"Q}(-i).
\end{equation}
As a consequence of the weak anisotropy, the polarization in an optically thick medium is low ($Q/I, U/I\ll 1$), and the anisotropy of the radiation field
is determined by the center-to-limb variation of the intensity $J^2_0=\oint\frac{{\rm d}\Omega}{4\pi}\frac{1}{2\sqrt{2}}[(3\mu^2-1)I_{\mu\chi'}$.
$J^2_0$ can be estimated from the emergent intensity \citep[e.g.,][]{MansoLandi02, MansoLandiEtal06}. 
Thus, for example, assuming a linear limb-darkening law for the intensity in the spectral line
\begin{equation}\label{eq05}
I=I^{(0)}(1-u+u\mu),
\end{equation}
where $\mu=\cos\theta$ and $u$ the limb-darkening coefficient, which is known from observations or numerical modeling \citep[e.g.,][]{PierceSlaughter77,Allende+04,Quirrenbach96, Nordgren+99,vanHamme93,Claret00}, or otherwise, can be left as a free parameter.
For the darkening law in Equation~(\ref{eq05}), $[J^0_0]_v=I^{(0)}(1-u/2)/2$, and $[J^2_0]_v=I^{(0)}u/16\sqrt{2}$.

For a dipole, the distribution of magnetic fields on the stellar surface is 
\begin{equation}\label{eqd}
{\boldsymbol B}=-\frac{B_d}{2}[{\boldsymbol e}-3({\boldsymbol e}\cdot{\boldsymbol r}){\boldsymbol r}],
\end{equation}
where $B_d$ is the magnetic field strength at the poles, and ${\boldsymbol e}$ is the unit vector directed along the dipole. 
For a point $P$ over the visible disk whose position is indicated by the unit vector ${\boldsymbol r}$ (see Figure 1b), Equations~(\ref{eq07}) can be written as
\begin{align}
[1+\delta^{(K)}(1-\epsilon)]\frac{2h\nu^3}{c^2}\frac{2J_\ell +1}{\sqrt{2J_u+1}} \rho^2_0&=
(1-\epsilon)w^{(2)}_{J_uJ_\ell} [\bar{J}^2_0]_v f_P(i, \Gamma), \label{eq08}
\displaybreak[0] \\
[1+\delta^{(K)}(1-\epsilon)]\frac{2h\nu^3}{c^2}\frac{2J_\ell +1}{\sqrt{2J_u+1}} \rho^2_{\pm 2}&=
(1-\epsilon)w^{(2)}_{J_uJ_\ell} [\bar{J}^2_0]_v [g_P(i, \Gamma)\pm {\rm i} h_P(i, \Gamma)],  \label{eq09}
\end{align}
where we have introduced the following functions of the dipole inclination $i$, strength $\Gamma_d$, at a given point $P$ over the stellar disk:
\begin{align}
f_P(i, \Gamma_d)&=
\sum_{Q'Q''}\frac{1}{1+\Gamma^2 Q'^2}
d^2_{0Q'}(\theta_B)d^2_{Q'Q''}(-\Theta_B)d^2_{Q"0}(-i)[\cos(Q''\ell)-\Gamma Q' \sin(Q''\ell)],      \label{eq10}
\displaybreak[0] \\
g_P(i, \Gamma_d)&=
\sum_{Q'Q''}\frac{1}{1+\Gamma^2 Q'^2}
d^2_{0Q'}(\theta_B)d^2_{Q'Q''}(-\Theta_B)d^2_{Q"2}(-i)[\cos(Q''\ell)-\Gamma Q' \sin(Q''\ell)],      \label{eq11}
\displaybreak[0] \\
h_P(i, \Gamma_d)&=
-\sum_{Q'Q''}\frac{1}{1+\Gamma^2 Q'^2}
d^2_{0Q'}(\theta_B)d^2_{Q'Q''}(-\Theta_B)d^2_{Q"2}(-i)[\Gamma Q'\cos(Q''\ell)+ \sin(Q''\ell)].      \label{eq12}
\end{align}
In Equations~(\ref{eq10})-(\ref{eq12}), the dependence on $P$ (and also on $i$), appear implicitly through (see Figure 1a-c) 
$$\Gamma^2=\Gamma_d^2 \frac{1}{4}[1+3({\boldsymbol e}\cdot{\boldsymbol r})^2],  \qquad
\cos\lambda={\boldsymbol e}\cdot{\boldsymbol r},   \qquad   \cos\theta_B=\frac{2{\boldsymbol e}\cdot{\boldsymbol r}}{\sqrt{1+3({\boldsymbol e}\cdot{\boldsymbol r})^2}}, \qquad
\cos \ell=\sqrt{1-\frac{[({\boldsymbol\Omega}\times{\boldsymbol e})\cdot{\boldsymbol r}]^2}{1-({\boldsymbol e}\cdot{\boldsymbol r})^2}},$$
where $\boldsymbol{\Omega}$ is the unit vector directed along the LOS.
In the saturation regime ($\Gamma_d\rightarrow\infty$), only the $Q'=0$ terms remain and
\begin{align}
f_P(i, \infty)&=
d^2_{00}(\theta_B)\sum_{Q"}d^2_{0Q"}(-\Theta_B)d^2_{Q"0}(-i)\cos(Q"\ell), \\
g_P(i, \infty)&=
d^2_{00}(\theta_B)\sum_{Q"}d^2_{0Q"}(-\Theta_B)d^2_{Q"2}(-i)\cos(Q"\ell), \\
h_P(i, \infty)&=-
d^2_{00}(\theta_B)\sum_{Q"}d^2_{0Q"}(-\Theta_B)d^2_{Q"2}(-i)\sin(Q"\ell).
\end{align}

Equations~(\ref{eq08})-(\ref{eq09}) are expressed in a common reference system ---with the quantization axis along the LOS---, and they can be averaged over the visible disk 
\begin{align}
[1+\delta^{(K)}(1-\epsilon)]\frac{2h\nu^3}{c^2}\frac{2J_\ell +1}{\sqrt{2J_u+1}} \rho^2_0&=
(1-\epsilon)w^{(2)}_{J_uJ_\ell} [\bar{J}^2_0]_v \bar{f}(i, \Gamma), \label{eq19}
\displaybreak[0] \\
[1+\delta^{(K)}(1-\epsilon)]\frac{2h\nu^3}{c^2}\frac{2J_\ell +1}{\sqrt{2J_u+1}} \rho^2_{\pm 2}&=
(1-\epsilon)w^{(2)}_{J_uJ_\ell} [\bar{J}^2_0]_v [\bar{g}(i, \Gamma)\pm {\rm i} \bar{h}(i, \Gamma)],  \label{eq20}
\end{align}
where we have introduced the following functions of the dipole inclination $i$ and
strength $\Gamma_d$:
\begin{align}
\bar{f}(i, \Gamma_d)&=\frac{1}{R_*^2}\iint {\rm d}\Sigma   
\sum_{Q'Q''}\frac{1}{1+\Gamma^2 Q'^2}
d^2_{0Q'}(\theta_B)d^2_{Q'Q''}(-\Theta_B)d^2_{Q"0}(-i)\cos(Q''\ell),      \label{eq21}
\displaybreak[0] \\
\bar{g}(i, \Gamma_d)&=\frac{1}{R_*^2} \iint {\rm d}\Sigma   
\sum_{Q'Q''}\frac{1}{1+\Gamma^2 Q'^2}
d^2_{0Q'}(\theta_B)d^2_{Q'Q''}(-\Theta_B)d^2_{Q"2}(-i)\cos(Q''\ell),      \label{eq22}
\displaybreak[0] \\
\bar{h}(i, \Gamma_d)&=\frac{1}{R_*^2} \iint {\rm d}\Sigma   
\sum_{Q'Q''}\frac{-\Gamma Q'}{1+\Gamma^2 Q'^2}
d^2_{0Q'}(\theta_B)d^2_{Q'Q''}(-\Theta_B)d^2_{Q"2}(-i)\cos(Q''\ell).      \label{eq23}
\end{align}
In Equations~(\ref{eq21})-(\ref{eq23}), the integral extends over the projected disk (see Figure 1c) and $R_*$ is the stellar radius.
In passing from Equations~(\ref{eq10})-(\ref{eq12}) to Equations~(\ref{eq21})-(\ref{eq23}), we have taken into account that two points at the same colatitude $\lambda$, and longitudes $\pm\ell$, have the same values of magnetic field strength ($\Gamma$) and inclination ($\theta_B$) ---hence, the same $\Theta_B$ too. 
Therefore, averages with $\sin(Q''\ell)$ cancel out.

The functions $\bar{f}(i, \Gamma_d)$, $\bar{g}(i, \Gamma_d)$, and $\bar{h}(i, \Gamma_d)$ are evaluated numerically and are shown in Figure~2. 
By construction they do not depend on the azimuth of the dipole with respect to the line of sight.
From this fact and the properties of the $d^2_{QQ'}$ rotation matrices in Equations~(\ref{eq10})-(\ref{eq12}), the
the following symmetries follow: $\bar{f}(i, \Gamma_d)=\bar{f}(\pi-i, \Gamma_d)$, $\bar{g}(i, \Gamma_d)=\bar{g}(\pi-i, \Gamma_d)$, $\bar{h}(i, \Gamma_d)=-\bar{h}(\pi-i, \Gamma_d)$.
It is also clear from symmetry arguments that in the absence of fields, the integrated linear polarization vanishes ---hence, $\bar{g}(i, 0)=\bar{h}(i, 0)=0$---, while $\bar{f}(i, 0)=\pi/4$\footnote{Clearly, $\bar{f}(i, 0)=\bar{f}(0, 0)$.
Now, making use of $d^2_{Q''0}(i=0)=\delta_{Q''0}$, and $\Sigma_{N}d^J_{MN}(\alpha)d^J_{NP}(\beta)=d^J_{MP}(\alpha+\beta)$ \citep[e.g.,][LL04]{BrinkSatchler68}, then $\bar{f}(0, 0)=\int\int d\Sigma d^2_{00}(-\lambda)=\pi/4$.}.
In the saturation regime ($\Gamma_d\rightarrow\infty$), $\bar{h}(i, \infty)=0$, and
\begin{align}
\bar{f}(i, \infty)&=\frac{1}{R_*^2} \iint {\rm d}\Sigma  \;  
d^2_{00}(\theta_B)\sum_{Q"}d^2_{0Q"}(-\Theta_B)d^2_{Q"0}(-i)\cos(Q"\ell), \\
\bar{g}(i, \infty)&=\frac{1}{R_*^2} \iint {\rm d}\Sigma  \;  
d^2_{00}(\theta_B)\sum_{Q"}d^2_{0Q"}(-\Theta_B)d^2_{Q"2}(-i)\cos(Q"\ell).
\end{align}
Consequently, in saturation, $\hat{\rho}^2_1=\hat{\rho}^2_2=0$, and $\epsilon_U=0$.
Light is thus polarized along the dipole and the maximum amount of linear polarization will be obtained at an inclination $i=90^\circ$.



\begin{figure}
\centerline{\includegraphics[width=10cm]{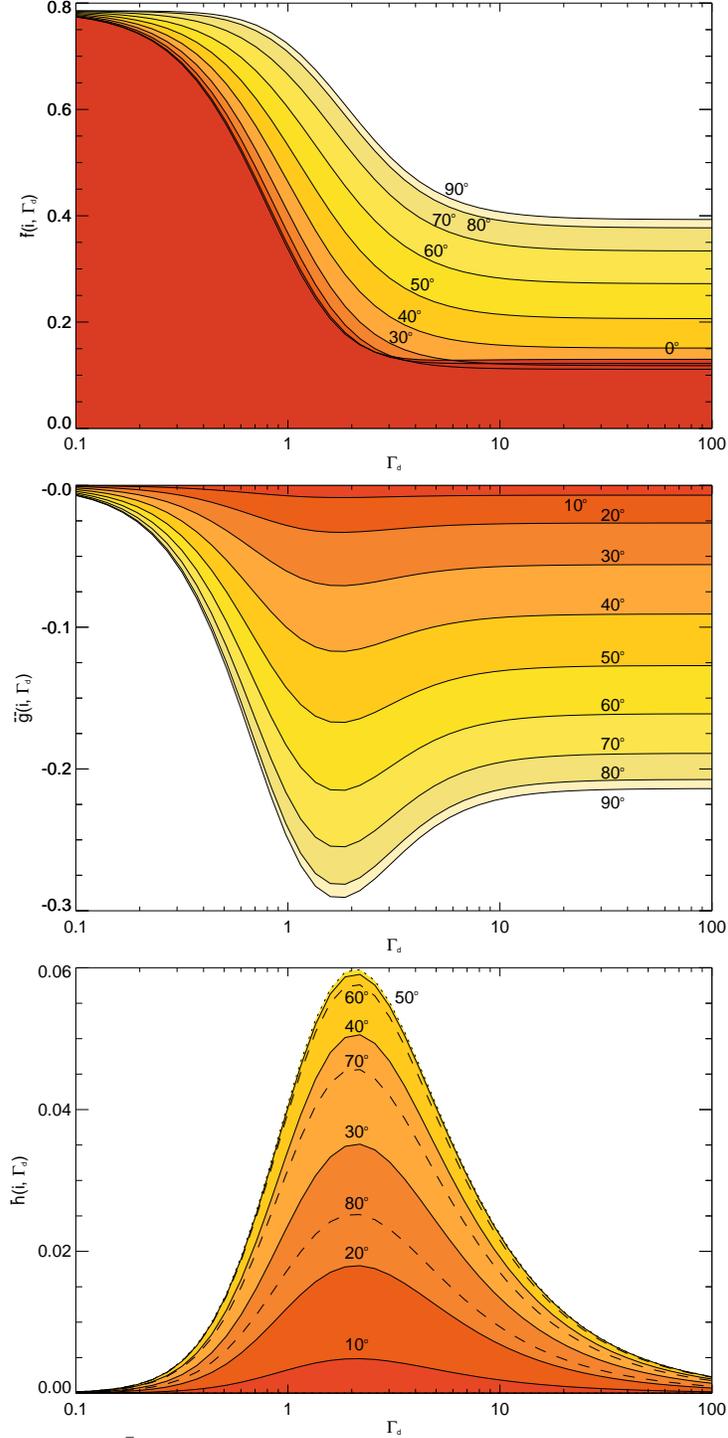}}
\caption{Functions $\bar{f}(i, \Gamma_d)$, $\bar{g}(i, \Gamma_d)$, and $\bar{h}(i, \Gamma_d)$, defined in Eqs.~(\ref{eq21})-(\ref{eq23}), for different values of the inclination angle $i$ of the dipole to the LOS. 
For $i>90^\circ$, $\bar{f}(i, \Gamma_d)=\bar{f}(180^\circ -i, \Gamma_d)$, 
$\bar{g}(i, \Gamma_d)=\bar{g}(180^\circ -i, \Gamma_d)$, and $\bar{h}(i, \Gamma_d)=-\bar{h}(180^\circ -i, \Gamma_d)$
Note that $\bar{f}(0, \Gamma_d)=\bar{f}(90^\circ, \Gamma_d)=\bar{g}(0, \Gamma_d)\equiv 0$), and that
$\bar{h}(i, \Gamma_d)$ reaches its maximum at $i\approx 52^\circ$.
\label{fig02}}
\end{figure}

The emergent Stokes parameters along the LOS ($z$-axis) from a semi-infinite atmosphere\footnote{The atmosphere cannot be considered semi-infinite very close to the extreme stellar limb. We neglect those effects here.} are 
\begin{gather}\label{eq15}
I=\int_0^{\infty}S_I\exp(-\tau){\rm d}\tau,   \qquad 
Q=\int_0^{\infty}S_Q\exp(-\tau){\rm d}\tau,   \qquad 
U=\int_0^{\infty}S_U\exp(-\tau){\rm d}\tau,   
\end{gather}
where $S_I=\epsilon_I/\eta_I$, $S_Q=\epsilon_Q/\eta_I$, $\epsilon_U/\eta_I$, 
$\eta_I=\eta_I^{\rm line}+\eta_I^{\rm cont}$, 
$\eta_I^{\rm line}=\frac{h\nu}{4\pi}B_{\ell u}{\cal N}_\ell\phi_\nu$ 
($B_{\ell u}$ is the Einstein coefficient for absorption), and
${\rm d}\tau=-\eta_I{\rm d}z$ is the element of optical depth,
which is measured from the stellar surface ($\tau=0$).
The integrals are approximated by the value of corresponding source functions $S_i$ at $\tau=1$,
which we evaluate from the emergent center-to-limb variation. 
The emergent fractional polarization then reads
\begin{gather}
\frac{Q}{I}=\frac{\epsilon_Q}{\epsilon_I}=\frac{\epsilon^{\rm line}_Q}{\epsilon^{\rm line}_I}
\frac{\alpha S_I^{\rm line}}{\alpha S_I^{\rm line}+B_\nu},  
\qquad\qquad
\frac{U}{I}=\frac{\epsilon_U}{\epsilon_I}=\frac{\epsilon^{\rm line}_U}{\epsilon^{\rm line}_I}
\frac{\alpha S_I^{\rm line}}{\alpha S_I^{\rm line}+B_\nu},  
  \label{eq16}
\end{gather}
where $\alpha=\eta_I^{\rm line}/\eta_I^{\rm cont}$, and $B_\nu=\epsilon_I^{\rm cont}/\eta_I^{\rm cont}$ and $S_I^{\rm line}=\epsilon_I^{\rm line}/\eta_I^{\rm line}$, are the source functions for the continuum (Planck function) and line, respectively. 
At the line core of a strong line, $\alpha\gg 1$ and $Q/I={\epsilon^{\rm line}_Q}/{\epsilon^{\rm line}_I}$, $U/I={\epsilon^{\rm line}_U}/{\epsilon^{\rm line}_I}$; 
for a weak line, $Q/I={\epsilon^{\rm line}_Q}/{\epsilon^{\rm line}_I}
\alpha\frac{S_I^{\rm line}}{B_\nu}(1-\alpha\frac{S_I^{\rm line}}{B_\nu}+...)$, 
$U/I={\epsilon^{\rm line}_U}/{\epsilon^{\rm line}_I}
\alpha\frac{S_I^{\rm line}}{B_\nu}(1-\alpha\frac{S_I^{\rm line}}{B_\nu}+...)$.

If the emissivities $\epsilon_i$ (see Equations~(\ref{eq00})-(\ref{eq0000})) are evaluated from Equation~(\ref{eq06}), then Equations~(\ref{eq16}) express the fractional polarization at a given point on the resolved stellar disk;
if they are evaluated from Equations~(\ref{eq08})-(\ref{eq12}), then we get the 
polarized fluxes ${\cal F}_I$, ${\cal F}_Q$, and ${\cal F}_U$, integrated over the stellar disk.
For symmetry reasons, $\epsilon_Q^{\rm cont}$ does not contribute to the integrated fractional polarization in Equation~(\ref{eq16}), though Rayleigh and Thompson scattering polarization should be included if the stellar disk were resolved.
If the reference direction for positive-$Q$ is taken along the dipole projected on the plane of the sky (i.e., $\chi=0$ in Equations~(\ref{eq000})-(\ref{eq0000})), then ${\cal F}_I$, ${\cal F}_Q$, and ${\cal F}_U$ are univocally determined by the functions $f$, $g$, and $h$, respectively. 
In particular, the symmetries discussed above for $f$, $g$, and $h$ apply to ${\cal F}_I$, ${\cal F}_Q$, and ${\cal F}_U$.
Finally, note that the not only the linear polarization, but also the total flux ${\cal F}_I$, is modulated by the presence of a global dipolar field on the stellar surface (through $f(i, \Gamma_d)$); this modulation is maximum for a dipole aligned with the LOS and minimum for a transversal dipole.

We note in passing, that the maximum amount of fractional polarization is obtained, in the absence of magnetic fields, at the stellar limb; from Equations~(\ref{eq00})-(\ref{eq06})
\begin{equation}
p_L^{\rm max}=\frac{\sqrt{Q^2+U^2}}{I}_{\rm max}=\frac{1}{2\sqrt{2}}\frac{3W_2[J^2_0]_v}{[1+\delta(1-\epsilon)][J^0_0+\frac{\epsilon}{1-\epsilon}B_\nu]-W_2[J^2_0]_v},
\end{equation}
where $W_2=(w_{J_uJ_\ell}^{(2)})^2$ is the line polarizability. 
If there is no limb-darkening or brightening, $u=0$, $[J^2_0]_v=0$, and hence, $p_L^{\rm max}=0$; if the radiation field at the surface is highly collimated ($[J^2_0]_v=J^0_0/\sqrt{2}$), and the atmosphere collisionless ($\epsilon=\delta^{(K)}=0$), the ideal limit of a 90$^\circ$ single-scattering event $p_L^{\rm max}=3W_2/(4-W_2)$ is recovered \citep[cf., Equation~(10.26) in][]{LandiLandolfi04}.

\section{Resolved dipolar disk}

\begin{figure}
\centering\includegraphics[width=14cm]{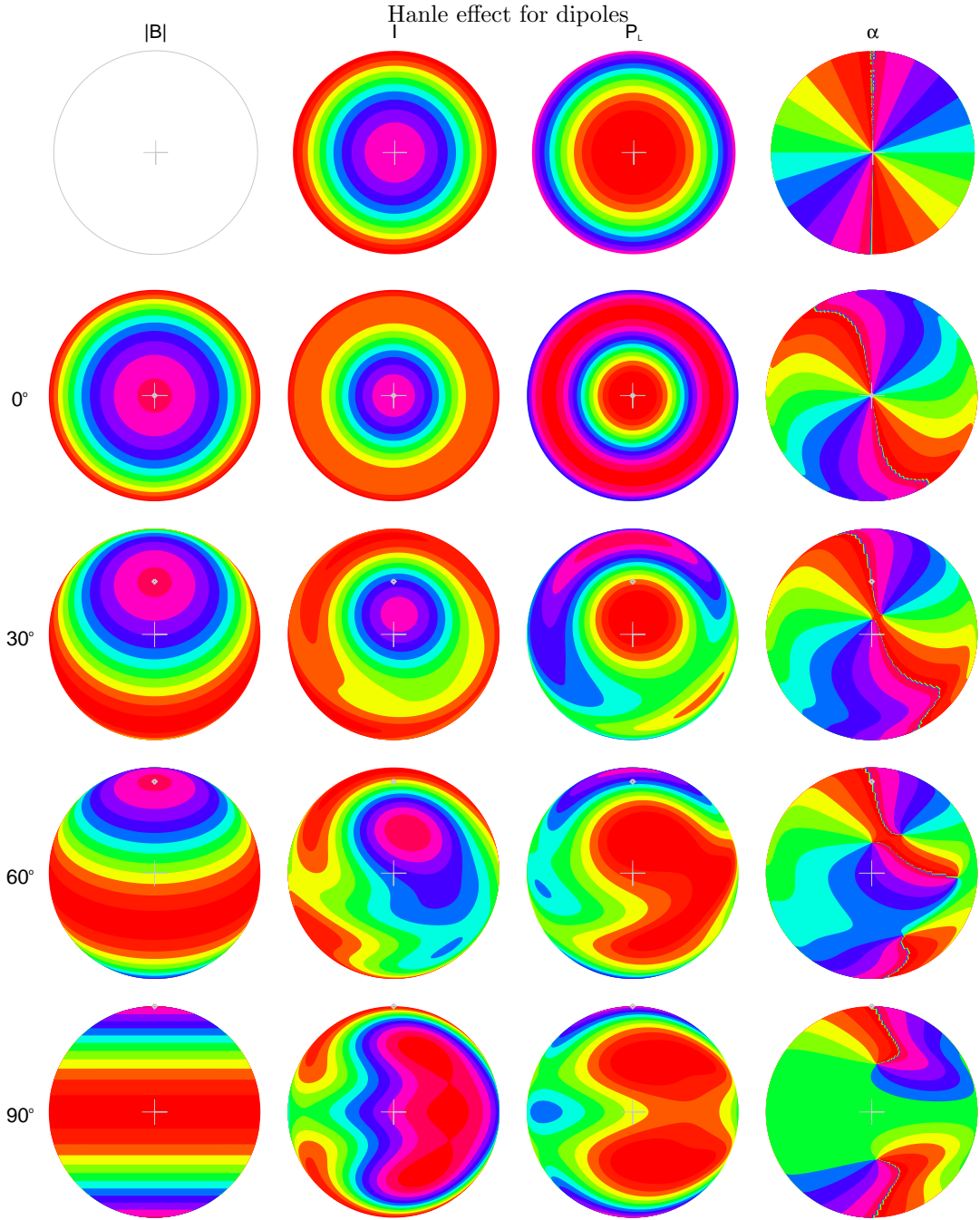}
\caption{Distribution of intensity and linear polarization over the disk of a dipole.
First column on the left shows, as a reference, the absolute value of the magnetic field strength $|B|$ ---which varies by a factor 2 between poles and equator. 
Columns 2 to 4 show the intensity, degree of linear polarization $P_L=\sqrt{Q^2+U^2}$, and polarization angle $\alpha_L$ ($\tan(2\alpha_L)=U/Q$), respectively.
Uppermost row shows the pure scattering, zero field case;
rows 2 to 5 correspond to $\Gamma_d=2$ and inclinations to the LOS $i=0^\circ$ (pole-on), $30^\circ, 60^\circ$, and $90^\circ$ (transversal), respectively. 
Crosses mark disk centers; diamonds, poles.
\label{fig03}}
\end{figure}

It is illustrative to study the distribution of intensity and polarization over the resolved disk of the dipole.
This problem may also be of practical interest for example, to study the emission of planets at long wavelengths where their intrinsic emission dominates over the irradiation of the Sun. 

Using Equations~(\ref{eq08})-(\ref{eq12}) into Equations~(\ref{eq00})-(\ref{eq0000}) and then, into Equations~(\ref{eq16}), we obtain the emergent fractional polarization at every point over the stellar disk (Figure~\ref{fig03}). 
We consider a strong line $\alpha=10^3$, with NLTE parameter $\epsilon=10^{-4}$.
In the absence of magnetic fields (upper most row in Figure~\ref{fig03}), the intensity shows the limb darkening law (Equation~(\ref{eq05}); second column of Figure~\ref{fig03}), the polarization increases from center (where it vanishes for clear symmetry reasons; third column of Figure~\ref{fig03}) to the limb ($p_{\rm max}$), and the polarization plane is always parallel to the stellar limb (see fourth column of Figure~\ref{fig03}).

Now, we consider a dipolar field with $\Gamma_d=2$, observed pole-on (i.e., inclined at an angle $i=0^\circ$ to the LOS; second row of Figure~\ref{fig03}). 
Due to the rotational symmetry, the disk center is still brighter but the limb darkening law is distorted with respect to the non-magnetic case due to the modification of the alignment $\rho^2_0$ in Equation~(\ref{eq00}) by the presence of magnetic fields.
The polarization is no longer maximal at the limb, but in a ring within the disk. This is because the magnetic field at the equator is pointing away from the LOS, a configuration that maximizes depolarization and rotation of the polarization plane characteristic of the Hanle effect. At disk center the magnetic field is vertical, and the Hanle effect does not operate (also clear from symmetry reasons). Half way between these to points, the polarization reaches a maximum, while the rotation of the polarization angle varies with the relative inclination of the magnetic field to the LOS, hence with the distance to disk center.

When the dipole is inclined $i=30^\circ$,  $60^\circ$, and $90^\circ$ (perpendicular) to the LOS (rows 3-5 of Figure~\ref{fig03}, respectively), the magnetic field at disk center becomes inclined to the LOS, and the symmetry arguments above do not apply anymore. 
Thus, the brighter region on the disk is displaced from the center, and limb darkening depends on the distance from the disk center as well as on the azimuth. Actually, in some cases, several bright regions may appear.
These {\em bright spots} are not due to abundance gradients (we are considering a homogeneous atmosphere) but just to a modulation of the intensity by the magnetic field, and if not properly accounted for, they could distort the interpretation of Doppler imaging techniques (see next section).
The behavior of polarization is likewise complex, depending on the distance to the limb but also on the azimuth.
Only in the perpendicular dipole configuration some symmetry is recovered, with the distribution of intensity and polarization being symmetric with respect to the equator.

Finally, it is interesting to observe the behavior of the singularities of the linear polarization field (points where the linear polarization vanishes and hence, the linear polarization angle is undefined) with the inclination of the dipole, which is most clearly illustrated in the rightmost panels of Figure~\ref{fig03}.
In the absence of magnetic fields or for a pole-on configuration, there is just one singular point for $\alpha_L$: disk center. With an inclination $i=30^\circ$ two close singularities appear slightly off-center, four at $60^\circ$ (two close ones in the north hemisphere, one in the south, and one close to the equator), and one at each hemisphere at $90^\circ$ inclination.
Now, consider a small closed path around a singular path and we follow it say, clockwise. 
For the first two cases (non-magnetic; pole-on dipole), we change twice color (i.e., polarization direction), and as we follow such a path we pass from blue to green to yellow. We arbitrarily assign to such a singularity an index +1. 
Proceeding along similar paths around the singularities for the $i=30^\circ$ inclination, we find the same color (polarization direction) change pattern (blue-green-yellow-...), but every color is found only once every turn. We assign such singularities an index +1/2. 
Three of the four singularities appearing in the $i=60^\circ$ inclination case follow this behavior too, and have index +1/2, but the fourth (the one close to the equator) changes once in reverse order (yellow-green-blue-...), for which an index -1/2 is assigned to it.
Finally, the two singularities in the $i=90^\circ$ case have indexes +1/2.
It is then easy to verify that the total amount of these indices is conserved ($+1=+1/2+1/2=+1/2+1/2+1/2-1/2$). This is closely related to the topology of singular points in polarization fields \citep{BerryDennisEtal04}.
It would be very interesting to consider the diagnostic value of such points. Just counting them or knowing their approximate location could be used to constrain the global topology and strength of the magnetic field.

\section{Oblique Rotators}

An oblique rotator is a star whose magnetic axis is inclined at an angle to the rotational axis. 
As a consequence, the polarimetric signal varies periodically as the star rotates.

\begin{figure}
\centering\includegraphics[width=18cm]{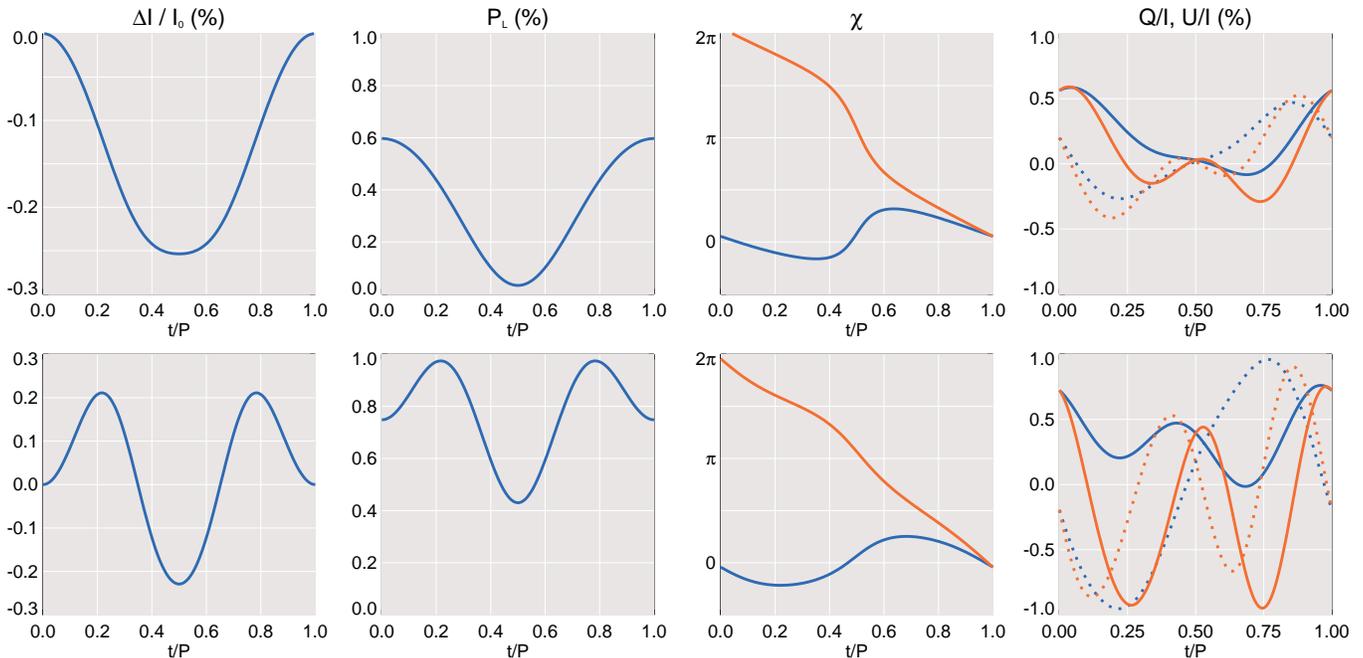}
\caption{Periodic behavior of the Hanle effect modulated polarized emission from an unresolved oblique rotator with a dipolar distribution of fields.
From left to right: intensity fluctuation normalized to the intensity at $t=0$, linear polarization $p_L$, azimuthal angle $\alpha$ ($\tan (2\alpha)={\cal F}_U/{\cal F}_Q$), and ${\cal F}_Q/{\cal F}̣_I$ (solid lines) and ${\cal F}_U/{\cal F}_I$ (dotted lines).
Upper and lower panels show two different general behaviors.
Upper panels: $i<90^\circ$ at all times (blue line: $j=30^\circ$, $\Psi=20^\circ$; red line: $j=20^\circ$, $\Psi=30^\circ$). Lower panels: $i$ fluctuates above and below $90^\circ$ at different times (blue line: $j=80^\circ$, $\Psi=40^\circ$; red line: $j=40^\circ$, $\Psi=80^\circ$). 
Note that the red and blue lines superpose in the two left columns.
\label{fig04}}
\end{figure}

Let $\Psi$ be the angle between the magnetic dipole ($\boldsymbol{e}$) and rotation ($\boldsymbol{\omega}$) axes, 
$j$ the inclination of the rotation axis to the LOS, and let's choose
the positive-$Q$ direction as the plane containing the rotation axis and LOS (see Figure 1d), then
\begin{align}
\cos i(f) &= -\sin\Psi\cos f \sin j+\cos\Psi\cos j, \label{eq27}\\
\sin \chi(f)\sin i(f) &= \sin \Psi\sin f, \label{eq28} \\
\cos \chi(f)\sin i(f) &= \sin\Psi\cos f \cos j+\cos\Psi\sin j, \label{eq29}
\end{align}
where $f$ ($0\le f\le 2\pi$) is the rotational phase angle ($f=0$ when the LOS, $\boldsymbol{\omega}$, and $\boldsymbol{e}$ are coplanar and thus ordered; see Figure 1d).
Using Equations~(\ref{eq27})-(\ref{eq29}) into the expressions (\ref{eq19})-(\ref{eq23}) and
(\ref{eq00})-(\ref{eq0000}), 
we calculate the intensity and polarization with rotational phase (Figure~\ref{fig04}).

The curves of brightness and polarization fluctuation fall in one of two broad classes according to their inclination of the dipole to the LOS being $i<90^\circ$ at all times during a rotation, or its changing from $i<90^\circ$ to  $i>90^\circ$. 
In the former case, the brightness and linear polarization vary once between maximum and minimum values in one period (upper panels in Figure~\ref{fig04}); in the latter, there are two maxima and two local minima (lower panels in Figure~\ref{fig04}).
Fluctuations in total brightness and polarization are symmetric with respect to the middle part of the rotational phase. 
This is because in every rotation, the two configurations of the dipole at $\pm f$ (or, equivalently, at $f$ and $2\pi-f$), show the same inclination $i$ to the LOS, their only difference being their azimuth, or equivalently, an inessential rotation of the reference system for the polarization direction.
Therefore, ${\cal F}_I(f)={\cal F}_I(2\pi-f)$, and $p_L(f)=p_L(2\pi-f)$ \citep[in contrast to][]{LopezAristeAsensioRamosEA11, IgnaceHoleEA11}.

There are two different behaviors for the azimuthal angle which is illustrated in the third column of Figure~(\ref{fig04}).

\begin{figure}
\centering\includegraphics[width=16cm]{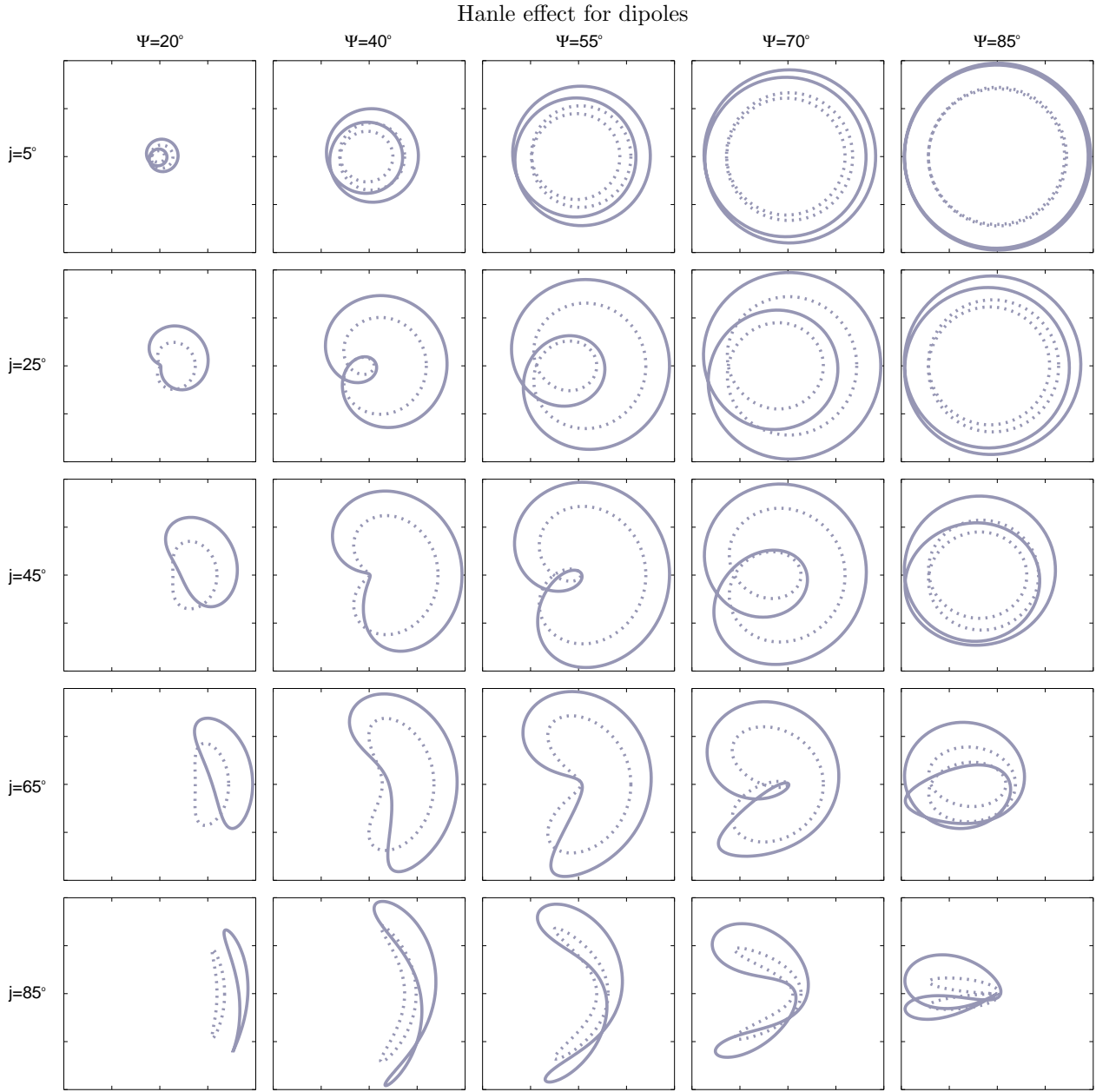}
\caption{Polarization diagrams ${\cal F}_U/{\cal F}_I$ versus ${\cal F}_Q/{\cal F}_I$ for a dipolar magnetic field configuration with $\Gamma_d=2$
(solid lines) and in saturation ($\Gamma_d=\infty$; dotted lines), at an angle $\Psi$ to the rotation axis, which in turn forms an angle $j$ with the LOS.
Each diagram corresponds to a full star rotation.
The diagrams show polarization varying between $\pm 1$\% (grid with steps of 0.5\%).
We consider $u=0.2$, $\epsilon=10^{-4}$, $\delta^{(K)}=0$.
\label{fig05}}
\end{figure}

Figure~\ref{fig05} shows polarization diagrams (${\cal F}_U/{\cal F}_I$ versus ${\cal F}_Q/{\cal F}_I$) along the rotation cycle, for different values of the $j$ and $\Psi$ angles. 
These diagrams are computed choosing the reference direction for positive-$Q$ along the projected rotation axis, and considering the only Hanle effect polarization. 
In actual observations, the orientation of the rotation axis on the plane of the sky forms an unknown angle $\varphi$ with respect to the arbitrary reference direction chosen for polarization. On the other hand, interstellar polarization is often not negligible, adding an arbitrary amount of linear polarization (constant along the line profile). 
The former effect amounts to a rotation by an angle $2\varphi$ of the diagrams in the $Q$-$U$ space; the latter to a translation. Therefore, the orientation and position of the diagrams may change, but their {\em shape} remains unchanged, being an intrinsic characteristic of the dipole.
Diagrams similar to these have been obtained for different mechanisms capable of generating linear polarization in spectral lines, like for example the differential saturation or magnetic intensification mechanism \citep{LandiEtAl81,LandolfiEtAl93}, although for more intense magnetic fields.

The degree of integrated linear polarization seen at $\pm f$ (or $f$ and $2\pi-f$) is the same, 
but the diagrams are not necessarily symmetric with respect to the ${\cal F}_U=0$ lane in general. 
In the saturation regime, however, the polarization signal depends on the direction of the magnetic field, but not on its orientation; hence, ${\cal F}_Q(f)={\cal F}_Q(-f)$, ${\cal F}_u(f)=-{\cal F}_u(-f)$, and the diagrams are symmetric with respect to ${\cal F}_U=0$.

\section{Stellar Quadrupoles}

We consider a distortion of the dipolar field in the form of an additional multipolar component of the global field. 
For the sake of simplicity, we consider a quadrupole aligned with the dipole ${\boldsymbol e}$.  
The contribution of the quadrupole to the distribution of magnetic fields on the stellar surface is 
\begin{equation}
{\boldsymbol B}_q=-\frac{B_q}{2} \{2({\boldsymbol e}\cdot{\boldsymbol r}){\boldsymbol e}+[1-5({\boldsymbol e}\cdot{\boldsymbol r})^2]{\boldsymbol r}\},
\end{equation}
where $B_q$ is the magnetic field strength at the poles.
The total magnetic field on the stellar surface is now ${\boldsymbol B}={\boldsymbol B}_d+{\boldsymbol B}_q$, where ${\boldsymbol B}_d$ is given by Equation~(\ref{eqd}).
The global field is still a poloidal field, and Equation~(\ref{eq07}) applies.
From it we may derive expressions for the resolved and integrated disk emission as in 
Equations~(\ref{eq08})-(\ref{eq09}) and (\ref{eq19})-(\ref{eq20}), respectively, but introducing
more general functions $f_P(i, \Gamma_d, \Gamma_q)$, $g_P(i, \Gamma_d, \Gamma_q)$, $h_P(i, \Gamma_d, \Gamma_q)$,
and the corresponding $\bar{f}(i, \Gamma_d, \Gamma_q)$, $\bar{g}(i, \Gamma_d, \Gamma_q)$, $\bar{h}(i, \Gamma_d, \Gamma_q)$, where $\Gamma_q$ is given from $B_q$ as after Equation~(\ref{eq02}).
The expressions of these functions are formally identical to their dipolar counterparts (Equations~(\ref{eq10})-(\ref{eq12}) and Equations~(\ref{eq21})-(\ref{eq23})), but now their dependence on the quadrupolar component is implicitly given by the normalized strength and inclination of the local magnetic field
\begin{gather*}
\Gamma^2=\frac{\Gamma_d^2}{4}[1+3({\boldsymbol e}\cdot{\boldsymbol r})^2] 
+ \frac{\Gamma_q^2}{4} [1-2({\boldsymbol e}\cdot{\boldsymbol r})^2+5({\boldsymbol e}\cdot{\boldsymbol r})^4]
+2\Gamma_d \Gamma_q ({\boldsymbol e}\cdot{\boldsymbol r})^3,  
\\
\cos\theta_B=\frac{2({\boldsymbol e}\cdot{\boldsymbol r})-q[1-3({\boldsymbol e}\cdot{\boldsymbol r})^2]]}{[(1+q^2)+(3-2q^2)({\boldsymbol e}\cdot{\boldsymbol r})^2+8q({\boldsymbol e}\cdot{\boldsymbol r})^3+5q^2({\boldsymbol e}\cdot{\boldsymbol r})^4]^{1/2}},
\end{gather*}
where $q=B_q/B_d$.

\begin{figure}
\centering\includegraphics[width=16cm]{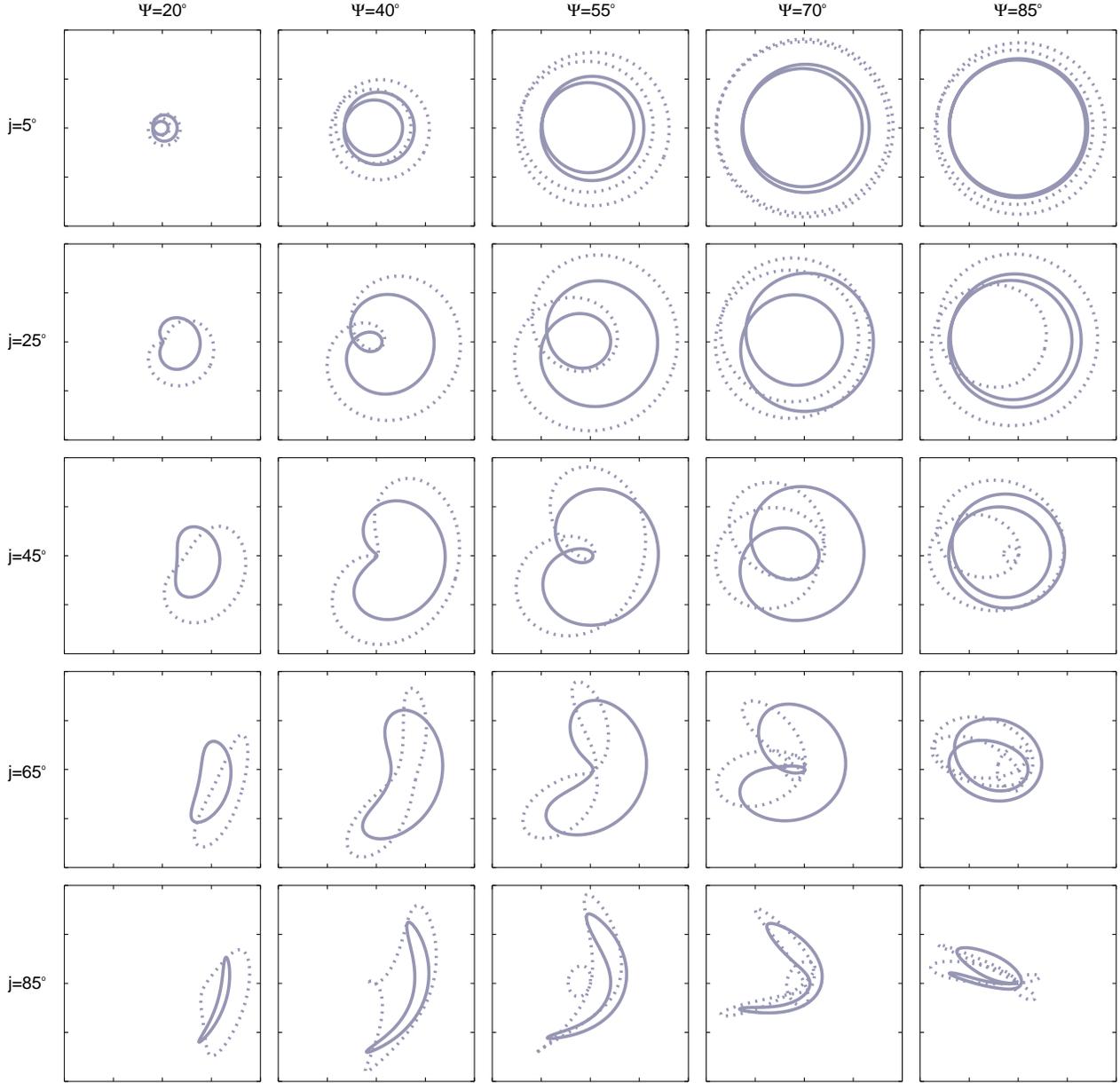}
\caption{Same as Figure~5 but for a purely quadrupolar distribution of fields ($q=\infty$, $\Gamma_q=2$; dotted lines), and a dipole+quadrupole ($q=1$, $\Gamma_d=2$; solid lines) distribution of magnetic fields.
\label{fig06}}
\end{figure}

Figure~\ref{fig06} shows polarization diagrams for a dipole+quadrupole configuration ($\Gamma_d=2$, $q=1$), and a pure quadrupole ($\Gamma_d=0$, $\Gamma_q=2$).
A purely quadrupolar distribution of fields tends to distort the polarization diagrams in an opposite sense to the dipolar distribution considered in Figure~\ref{fig05}, and consequently, the dipolar+quadrupolar distribution tends to be more symmetric with respect to the ${\cal F}_U=0$ line. This symmetrization and the fact that the total polarization in the latter case is also smaller is due to the fact that the strength of the field is everywhere larger. 
In the saturated regime, the diagrams should be symmetric about the ${\cal F}_U=0$ line, as explained above.
Note also, that in the present case, the polarization in the saturated regime depend on the parameter $q$.

All the above considerations can be extended to more general distributions of fields (octupolar, hexapolar, etc.), as long as the additional components are restricted to be aligned with the dipole and hence, the global field remains poloidal. 
For a general arbitrary orientation of the quadrupole (or larger multipoles) with respect to the dipole, a toroidal component of the field appears, and Equation~(\ref{eq07}) from which all our analysis derives is no longer applicable. 
A further generalization to include these more general geometries is, nonetheless, possible along similar lines to the method followed here.

\section{Conclusions}

We have calculated the polarized emission by an unresolved stellar dipole, and Equations (\ref{eq19})-(\ref{eq20}) are the main result of this paper.
To obtain them, we have derived a general expression (Equation~\ref{eq06}) and then made several approximations. 

The emissivity in a spectral line in a given direction is simply given from the atomic polarization of its upper level expressed in that ray reference system (Equations~(\ref{eq00})-(\ref{eq0000})).
The statistical equilibrium (balance between radiative processes, collisions, and magnetic precession) of the atomic polarization in the magnetic field reference system is given by Equation~(\ref{eq02}).
The symmetries of the radiation field are naturally expressed by the radiation field tensors $J^K_Q$ in the atmosphere's local vertical reference system.
By rotating from the local vertical to the local magnetic field system, then to the common dipole reference system, and finally to the LOS reference system, we get the exact Equation~(\ref{eq06}), giving the statistical tensors in a common reference system and which is then amenable to integration over the stellar disk for a poloidal distribution of magnetic fields.

We have made the following approximations.
1) Assuming that the radiation field is axially symmetric with respect to the radial direction, we can express the disk integrated atomic polarization (hence, scattering polarization) in terms of just the three real functions $f_P(i, \Gamma_d)$, $g_P(i, \Gamma_d)$, and $h_P(i, \Gamma_d)$, whose averages over the stellar disk are shown in Figure~2.
2) The mean intensity $J^0_0$ and the anisotropy $J^2_0$ may be calculated from the intensity in a given model atmosphere. 
Alternatively, here, we estimate their value at the stellar surface ($\tau=1$) from the emergent center-to-limb variation.
This is the strongest approximation in our approach. 
The actual values of the radiation field tensors can be derived from more detailed theoretical calculation or observations, or be left as a convenient parametrization to be constrained from observations (together with the magnetic field strength and orientation).

This approach allows us to derive explicit expressions for the distribution of intensity and polarization over a resolved disk and the behavior of the polarized flux for unresolved stellar dipoles.
The results can be applied to unresolved stellar dipoles, as well as to planetary dipoles (resolved or not) at long wavelengths where the emission of the planet dominates over the stellar illumination.

\end{document}